\documentclass[ALICE,manyauthors]{cernphprep}
\usepackage{cite, rotating, amssymb, epsfig, lineno, color}
\usepackage{amsmath}
\usepackage{hyperref}
\usepackage{xspace}
\usepackage{xcolor,soul}
\usepackage{xparse}
\usepackage{xstring}
\usepackage{euscript}
\usepackage[utf8]{inputenc}
\usepackage[english]{babel}
\usepackage{graphicx}
\usepackage{blindtext}
\usepackage{float}
\usepackage{adjustbox}
\usepackage{enumitem}
\usepackage{rotating}
\usepackage{graphicx}
\usepackage[T1]{fontenc}
\usepackage{orcidlink}

\newcommand{\fourpi}{$\pi^+\pi^-\pi^+\pi^-$\hspace*{0.1mm}}
\newcommand{\rhopipi}{$\rho^0\pi^+\pi^-$ \hspace*{0.1mm}}
\newcommand{\fivenn} {$\sqrt{s_{\mathrm{NN}}}~=~5.02$~Te\kern-.1emV}
\newcommand{\GeVc} {Ge\kern-.1emV/$c$} 
\newcommand{\MeVc} {Me\kern-.1emV/$c$} 
\newcommand{\TeV} {Te\kern-.1emV} 
\newcommand{\GeV} {Ge\kern-.1emV} 
\newcommand{\MeV} {Me\kern-.1emV} 
\newcommand{\GeVmass} {Ge\kern-.1emV/$c^2$} 
\newcommand{\MeVmass} {Me\kern-.1emV/$c^2$} 
\newcommand{\pt}{$p_\mathrm{T}$}

\begin{document}

%%%%%%%%%%%%%%%  Title page %%%%%%%%%%%%%%%%%%%%%%%%

\begin{titlepage}
\PHyear{2024}
\PHnumber{104}       % required, will be obtained from PH
\PHdate{10 April}   % required, will be obtained from PH

\title{Exclusive photoproduction of excited $\rho$ mesons decaying to four pions in ultraperipheral Pb--Pb collisions at 
$\mathbf{\sqrt{\textit{s}_{\textup{NN}}} = 5.02}$\;TeV}
\ShortTitle{Four pion photoproduction}

%%% Do not change the next lines
\Collaboration{ALICE Collaboration\thanks{See Appendix~\ref{app:collab} for the list of collaboration members}}
\ShortAuthor{ALICE Collaboration} % appears on left page headers, do not change

\begin{abstract}
\par The intense photon fluxes from relativistic nuclei provide an opportunity to study photonuclear interactions in ultraperipheral collisions.  In particular, it allows for the investigations of excited, light-flavour vector mesons. The measurement of coherently photoproduced \fourpi~final states in ultraperipheral Pb--Pb collisions at \fivenn~is presented for the first time. The cross section, d$\sigma$/d$y$, times the branching ratio ($\rho\rightarrow \pi^+ \pi^+ \pi^- \pi^-$) is found to be $47.8\pm2.3~\rm{(stat.)}\pm7.7~\rm{(syst.)}$~mb in the rapidity interval $|y| < 0.5$. The invariant mass distribution is not well described with a single Breit-Wigner resonance without an interference term. Including interference with a non-resonant contribution results in the mass and width values being too far from those reported in PDG, while the production of two interfering resonances, $\rho(1450)$ and $\rho(1700)$, also provides a good description of the data. The values of the masses ($m$) and widths ($\Gamma$) of the resonances extracted from the fit assuming two interfering resonances are $m_{1}=1385\pm14~\rm{(stat.)}\pm3~\rm{(syst.)}$~\MeVmass, $\Gamma_{1}=431\pm36~\rm{(stat.)}\pm82~\rm{(syst.)}$~\MeVmass, $m_{2}=1663\pm13~\rm{(stat.)}\pm22~\rm{(syst.)}$~\MeVmass~and $\Gamma_{2}=357~\pm31~\rm{(stat.)}\pm49~\rm{(syst.)}$ \MeVmass, respectively. The measured cross sections times the branching ratios are compared to recent theoretical predictions. 
\end{abstract}
\end{titlepage}

\setcounter{page}{2} %please do not remove this line

\section{Introduction}
Collisions involving ultrarelativistic heavy ions offer a rich area for research. The distance between the centers of two nuclei at the moment of their closest approach, called the impact parameter, is one of the most important characteristics of the interaction. When this distance exceeds the sum of the nuclear radii,  ultraperipheral collisions (UPCs) can occur~\cite{Baltz:2007kq, Contreras:2015dqa, Klein:2017nqo}. In this case, the charges of all $Z$ protons in a nucleus act coherently and the photon fluxes from each nucleus are enhanced by a factor of $Z^2$ compared to proton beams.
This enhancement, together with the high beam energies at the LHC, produces strong fluxes of high-energy photons. 
At the same time, the requirement of coherent emission from nuclei limits the photon virtuality to $Q^2 < (\hslash/R_{A})^2$. Photons with higher virtuality are strongly suppressed by the nuclear electromagnetic form factor~\cite{Klein:2019qfb, Klein:2020fmr, Schafer:2020bnm, Frankfurt:2022jns}.

\par Photon-induced reactions at the LHC include purely electromagnetic photon-photon processes and photon-nucleus interactions. The latter includes exclusive processes where the photon fluctuates to a bound ${\rm q\overline{q}}$ system, typically a vector meson (VM), which then scatters elastically off the nucleus. The total cross section for this process can be factorized into the photon flux and the cross section of the corresponding interaction. 

\par Elastic scattering of the VM can proceed either off the entire target nucleus (coherently), where the nucleus usually remains intact, or off only one of the nucleons
(incoherently), where the target nucleus typically breaks up, emitting nucleons at very forward rapidities. For coherent processes, the size of the lead ion restricts the transverse momentum (\pt) of the VM to about 100~\MeVc, while it is below 1~\GeVc~for incoherent processes. VM photoproduction dominates the hadronic structure of the photon, and the main contribution to the total exclusive photoproduction cross section comes from the \mbox{$\gamma$~+~p $\rightarrow \rho^0$~+~p} process. Scaling from a proton to a nuclear target is often implemented using a Glauber approach assuming the VM dominance model~\cite{Bauer:1977iq}.

\par Coherent $\rho^0$ photonuclear production was thoroughly investigated in Au--Au UPCs at RHIC~\cite{STAR:2007elq, STAR:2002caw} and in Pb--Pb~\cite{ALICE:2020ugp}, p--Pb~\cite{CMS:2019awk} and Xe--Xe UPCs ~\cite{ALICE:2021jnv} at LHC. The cross section for this process and d$\sigma/$d$y$ distributions were found to be well described by a Glauber calculation. The study of the excited states of the $\rho^0$ meson is particularly intriguing. These may not just be composed of radial excitations of the $\rho^0$ but could also include hybrid (${\rm q\overline{q}g}$) states\cite{Barnes:1996ff,Close:1997dj}. The Particle Data Group (PDG)~\cite{ParticleDataGroup:2022pth} identifies $\rho(1450)$ and $\rho(1700)$ as at least two possible excited states based on the previous experimental measurements conducted mainly in $e^+e^-$ annihilation and at lower collision energies~\cite{L3:2005dor, Atiya:1979ip, Donnachie:1986pk, Schacht:1974fq, Davier:1973fy, DM2:1988uai, DM2:1988xqd, STAR:2009giy, WA76:1989mye, WA102:1997sum}. However, such data are relatively sparse and have large uncertainties. In UPCs, such a high-mass resonance was observed by the ALICE Collaboration~\cite{ALICE:2020ugp} in exclusive two-pion events. Another possible decay mode of an excited $\rho$ resonance involves four charged pions, \fourpi, in the final state. The photoproduction of the four-pion final state has been measured by the OMEGA spectrometer~\cite{OmegaPhoton:1984rzb, WA91:1994tzu, WA91:1995ffz} and in UPCs, by the STAR Collaboration at RHIC~\cite{STAR:2009giy}. Previous publications~\cite{L3:2005dor, Atiya:1979ip, Donnachie:1986pk, Schacht:1974fq, Davier:1973fy, DM2:1988uai, DM2:1988xqd, STAR:2009giy, WA76:1989mye, WA102:1997sum} suggested that the measured invariant mass spectrum may be attributed to the two aforementioned resonances. Nonetheless, the accuracy of the data was insufficient for distinguishing these two resonances and determining their mixing angle. In contrast, a recent preliminary measurement by the H1 Collaboration~\cite{H1:2018-011} suggests the possibility of fitting the data with a single broad resonance. To date, no measurements of this final state have been carried out at the LHC, making searches in the \fourpi final state crucial for understanding the nature of these resonances.

Finally, improved measurements of the exclusive photoproduction cross section of excited $\rho$ mesons have implications for the understanding of nuclear shadowing. Exclusive photoproduction of a $\rho^0$ off a nucleon ($\gamma N \rightarrow \rho^{0}N$) can proceed via the fluctuation of the photon to a $\rho^0$ meson which scatters elastically off the target. There is, however, also the possibility to have contributions from cross terms, where the fluctuation is to an excited $\rho$ which converts to a $\rho^0$ in the scattering process ($\rho N \rightarrow \rho^{0}N$). The importance of these cross terms for the interpretation of exclusive vector meson production in terms of nuclear shadowing has been pointed out earlier~\cite{Pautz:1997eh}.

\par This article reports on the first measurement of exclusive \fourpi~photoproduction in Pb--Pb UPCs at \fivenn. A resonance structure is found in the invariant mass spectrum and the cross section times branching ratio ($\rho\rightarrow \pi^+ \pi^+ \pi^- \pi^-$) is measured for an excited $\rho$ state in the rapidity interval $|y| < 0.5$. The possibility of two excited resonances and their mixing is also studied.

\section{Experimental setup}
The analyzed data were recorded by the ALICE Collaboration in the fall of 2015 when the LHC provided Pb--Pb collisions at \fivenn. A detailed description of the ALICE systems and their performance is given in Refs.~\cite{ALICE:2008ngc, ALICE:2014sbx}. The \fourpi~final state is reconstructed using the Inner Tracking System (ITS)~\cite{ALICE:2010tia} and the Time Projection Chamber (TPC)~\cite{Alme:2010ke} to measure the pion tracks. 
The Silicon Pixel Detector (SPD) makes up the first two layers of the ITS, the closest to the beam, and is used for both tracking and triggering purposes. 

The TPC is the main tracking detector. It is a large cylindrical gaseous detector with a central membrane at high voltage and readout planes, composed of multiwire proportional chambers at each of the two end caps. It covers the full azimuthal range and pseudorapidity interval $|\eta|<0.9$ for tracks which fully traverse it. 
The ITS and TPC are located inside a large solenoid magnet, creating a uniform 0.5~T magnetic field parallel to the beam-direction axis.

The V0 and AD detectors are used as veto detectors to reject hadronic events. The V0 detector~\cite{ALICE:2013axi} is a set of two segmented scintillators, V0A and V0C. V0A covers the  $2.8<\eta<5.1$ range, while V0C covers $-3.7<\eta<-1.7$.  The AD~\cite{LHCForwardPhysicsWorkingGroup:2016ote} detector is a set of two arrays of scintillation detectors, ADA and ADC, placed further away from the nominal interaction point and covering $4.7<\eta<6.3$ and $-6.9<\eta<-4.9$, respectively.
\par The trigger used to obtain the data sample for the measurement described in this Letter uses five signals: a topological SPD trigger and four vetoes of any activity within the time windows corresponding to nominal beam–beam interactions in ADA, ADC, V0A and V0C. SPD provides a topological trigger requiring at least two tracklets having an opening angle in azimuth larger than 153 degrees. Such a trigger leaves a sample with events containing at least two back-to-back tracks in central detectors.

\par The integrated luminosity is determined using a reference trigger based on the multiplicity of the V0A and V0C detectors~\cite{ALICE:2022xir}. The integrated luminosity of the analyzed sample was $622\pm16$~mb$^{-1}$. 

For signal extraction and corrections for acceptance and efficiency, two event samples, for both coherent and incoherent photoproduction, were generated using the STARlight Monte Carlo~\cite{Klein:2016yzr}, based on a Glauber-like eikonal formalism. The simulated excited $\rho$ meson events were generated according to a Breit-Wigner (B--W) shape discussed below with the mass and width equal to $m = 1720$~\MeVmass~and \mbox{$\Gamma=249$~\MeVmass}~and processed using realistic simulations of the ALICE detector based on GEANT 3.21~\cite{Brun:1994aa}.

\section{Event selection and background subtraction}
Events that meet the trigger criteria described above were selected if they contain exactly four good-quality tracks. The selected tracks were required to be reconstructed in both the ITS and the TPC with at least two hits in the ITS and 50 out of 159 space points in the TPC, or to be reconstructed only in the ITS with at least three hits. It was found that less than 1\% of four track events have tracks that do not satisfy the pion PID hypothesis. Thus, taking into account the limited statistics of our sample, it was decided to omit the PID selection and consider all tracks to be pions. In addition, each track was required to have a distance of closest approach to the event interaction vertex of less than 3.2~cm in the beam direction, and less than 2.4~cm in the plane transverse to the beam direction. All four tracks were required to have at least one matching hit seen by the trigger. The four-track events with zero net charge were used to create the signal sample, while non-zero net charge events were used to estimate the combinatorial background.

Since coherently produced VMs typically have small summed \pt~of the four constituent tracks (event \pt), events with a large \pt~are expected to be dominated by combinatorial background with more particles in the final state. However, given only four charged pions are detected, a good estimate of its \pt~spectrum can be obtained from non-zero net charge events.

\par
\begin{figure}[!th]
\includegraphics[width=1.\linewidth]{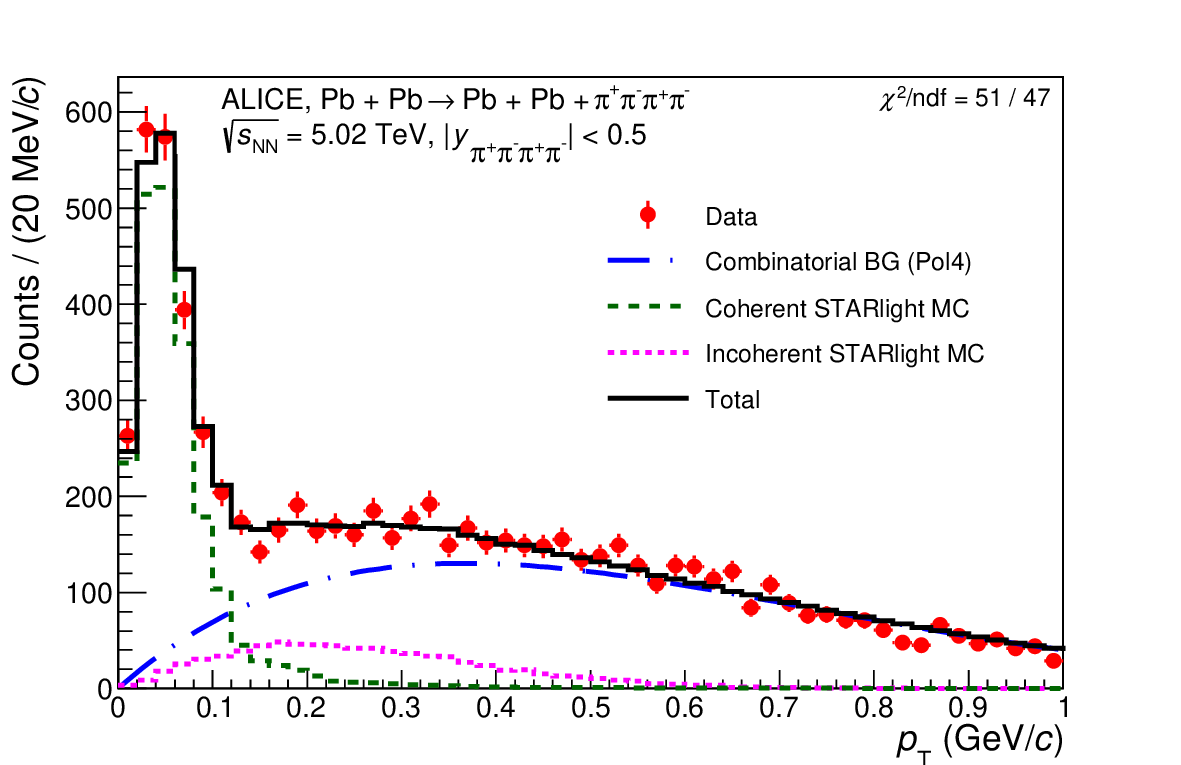}
\caption{\label{template} Distribution of the event \pt~in events with zero net charge in the invariant mass range \mbox{$0.8 < m < 2.5$~\GeVmass} and the rapidity interval $|y| < 0.5$. The data (red circles) are shown together with the fitted function (solid black line) and its three components as described in the text.}
\end{figure}

 In order to estimate the combinatorial and incoherent backgrounds in the signal region (\pt~$< 150$~\MeVc), the event \pt~distribution, in the invariant mass range $0.8 < m < 2.5$~\GeVmass, is fitted using a combination of three templates (Fig.~\ref{template}). First, the template corresponding to the combinatorial background (solid blue line) was obtained by fitting a fourth-order polynomial function to the \pt~distribution of non-zero net charge events with exactly four charged pions. The other two templates were obtained from STARlight MC samples for the coherent (dashed orange histogram) and incoherent (fine-dashed magenta histogram) processes, respectively. A $\chi^2$ minimization procedure was used to fit the data (red circles) to the sum of these three templates, where the normalization of each template was a free parameter. Since in the STARlight MC the \pt spectrum is calculated using a nuclear form factor rather than through a full Glauber calculation, it does not describe the position of the coherent peak precisely. Thus, the following iterative procedure was implemented. After the initial fit is performed, the combinatorial and incoherent contributions are subtracted bin-by-bin from the data sample. The ratio of the resulting spectra to the coherent MC sample follows a linear function in \pt, which was used to calculate the weight applied to the coherent MC sample on an event-by-event basis before the template fit was repeated. The final result of this procedure is shown in Fig.~\ref{template} as a solid black histogram for the sum of three templates. The normalization factor of the combinatorial background was found from the \pt$ > 600$~\MeVc~interval to be $1.60\pm0.15~\rm{(stat.)}$, resulting in the yields of the three contributions to the signal region (\pt~$<$ 150~\MeVc): $N_{\rm coh} = 1987~\pm~54~\rm{(stat.)}$, $N_{\rm incoh} = 134~\pm~13~\rm{(stat.)}$ and $N_{\rm comb} = 320~\pm~23~\rm{(stat.)}$, respectively. 

\section{Signal extraction}

The $m_{\pi^+\pi^-\pi^+\pi^-}$-dependent product of the cross section times the branching ratio was obtained as follows. First, the combinatorial background contribution, as obtained from the template fit shown in Fig.~\ref{template}, was subtracted from the invariant mass spectra of zero net charge events in the signal region \mbox{(\pt~$<$ 150~\MeVc)}. Second, this background-subtracted invariant mass spectrum was corrected bin-by-bin for detector acceptance and trigger efficiency ($A\times\varepsilon$). The $A\times\varepsilon$ correction factor was calculated using the coherent sample of events generated by STARlight~\cite{Klein:2016yzr}, which were processed through the ALICE detector response simulation~\cite{Brun:1994aa}. The correction was found to rise linearly as a function of event mass from 0 at 1.0~\GeVmass, reaching almost a constant value of $0.07$ around 1.5~\GeVmass. Finally, the corrected spectrum was scaled according to the following formula:
\begin{equation}
    \frac{{\mathrm d^2}\sigma}{{\mathrm d} y {\mathrm d} m }\times \mathrm{BR} =\frac{ N_{\rho}  \times (1-f_{\mathrm I}) \times f_{\mathrm p} }{ \Delta y \times \Delta m \times \mathcal{L}_{\text{int}} }.
\end{equation}
Here $\mathrm{BR}$ is the branching ratio ($\rho\rightarrow \pi^+ \pi^+ \pi^- \pi^-$), $N_{\rho}$ is the number of \fourpi events in each invariant mass bin after the acceptance and efficiency correction, $f_{\mathrm I}$ is the correction for the remaining incoherent contribution, equal to $(6.3\pm1.0~\rm{(stat.)})\%$, $\mathcal{L}_{\text{int}}$ is the integrated luminosity of the analyzed sample, equal to $622\pm16$~mb$^{-1}$, $\Delta y$ is the rapidity interval width in which the measurement is performed, equal to one unit of rapidity, and $\Delta m$ is the invariant mass bin width. Finally, $f_{\mathrm p}$ is the pileup correction, mainly from events with exchange of multiple photons that would invalidate the veto requirement in the dedicated detectors. The probability of pileup is correlated with the average number of inelastic hadronic collisions per bunch crossing. The pileup effect is estimated using two different methods described in detail in Ref.~\cite{ALICE:2020ugp}, and its value is $1.071\pm0.038$ in the current measurement, with the uncertainty taken as the difference between the two methods.

\par The amplitude of the resonance production is usually described by a relativistic Breit--Wigner (B--W) function as derived by Jackson
\cite{Jackson:1964zd}:
\begin{equation}
	BW_\mathrm{part} = \frac{\sqrt{m_\mathrm{part} \cdot m_\mathrm{event} \cdot \Gamma_\mathrm{event}}}
                  {m^2_\mathrm{event} - m^2_\mathrm{part} + i\cdot m_\mathrm{part} \cdot \Gamma_\mathrm{event}},
	\label{SoedingBW}
\end{equation}
where the mass-dependent width $\Gamma_\mathrm{event}$ is given by
\begin{equation}
	\Gamma_\mathrm{event} = \Gamma_\mathrm{part} \cdot \frac{m_\mathrm{part}}{m_\mathrm{event}} \cdot 
         \left( \frac{m^2_\mathrm{event}-k\cdot m^2_{\pi}}{m^2_\mathrm{part}- k\cdot m^2_{\pi}} \right)^{3/2}.
	\label{SoedingWidth}
\end{equation}
Here, $m_\mathrm{part}$, $\Gamma_\mathrm{part}$ are the mass and the width of a resonance, $m_\mathrm{event}$
is the event mass. The constant $k$ is equal to 4 for two-pion decays, since both pions have the same energy in the center-of-mass frame. For a four-pion decay, the value of $k$ is not well defined. If all four pions have the same energy in the center of mass, one finds $k = 16$. This is the most natural choice here, since this expression reflects how much the energy in the center of mass differs from the sum of the masses of the daughter particles. At the same time, it was found that the fit behavior and the extracted parameters do not show a strong dependence on the particular choice of the $k$ value.

\par In the present analysis, three approaches using the relativistic B--W formula are used to describe the data. The first considers a single resonance:
\begin{equation}
        \frac{{\mathrm d}\sigma}{{\mathrm d}m} = |A\cdot BW_{1}|^2,
        \label{SR}
\end{equation}
The second one is a single BW-resonance interfering with a constant non-resonant term. This was proposed by S\"oding~\cite{Soeding:1966zz}, and has been found to give a good description of exclusive $\rho^0$ production~\cite{STAR:2002caw, STAR:2007elq, ALICE:2020ugp, ALICE:2021jnv, CMS:2019awk}:
\begin{equation}
\frac{{\mathrm d}\sigma}{{\mathrm d}m} = |A\cdot BW_{1} + e^{-i\varphi}\cdot B|^2,
        \label{SRinter}
\end{equation}
and the final fit is to two interfering BW-resonances:
\begin{equation}
\frac{{\mathrm d}\sigma}{{\mathrm d}m} = |A\cdot BW_{1} + e^{-i\varphi}\cdot B\cdot BW_{2}|^2,
        \label{TRinter}
\end{equation}
where $A$ and $B$ are the normalization factors and $\varphi$ is the phase difference parameter. In addition to these, the masses and widths of the resonances are free parameters in the fit.

\begin{figure}[!h]
\includegraphics[width=.65\linewidth]{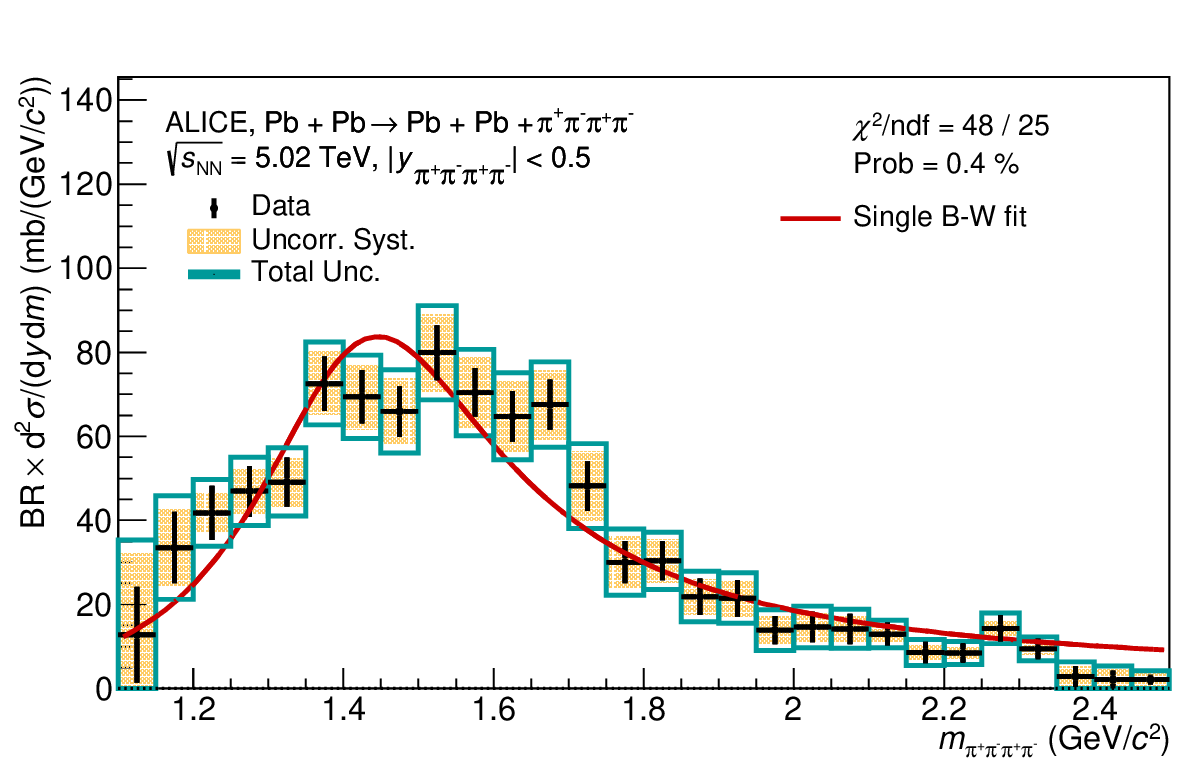}

\includegraphics[width=.65\linewidth]{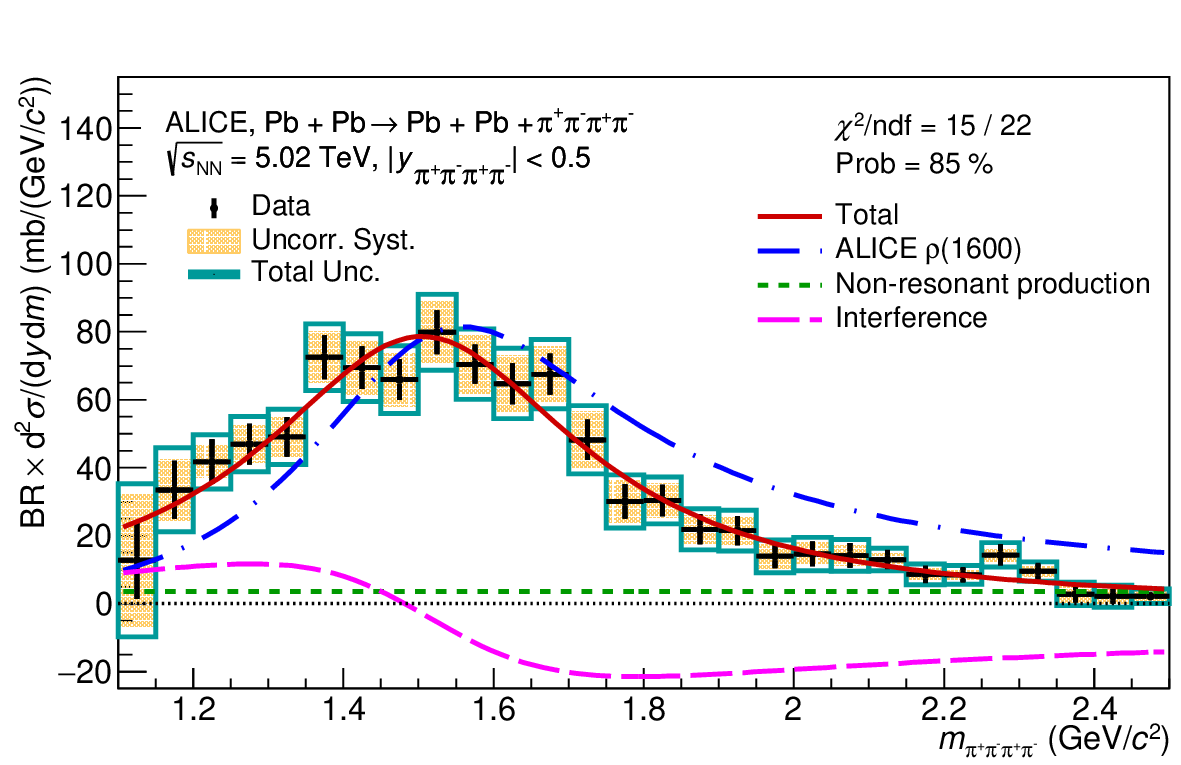}

\includegraphics[width=.65\linewidth]{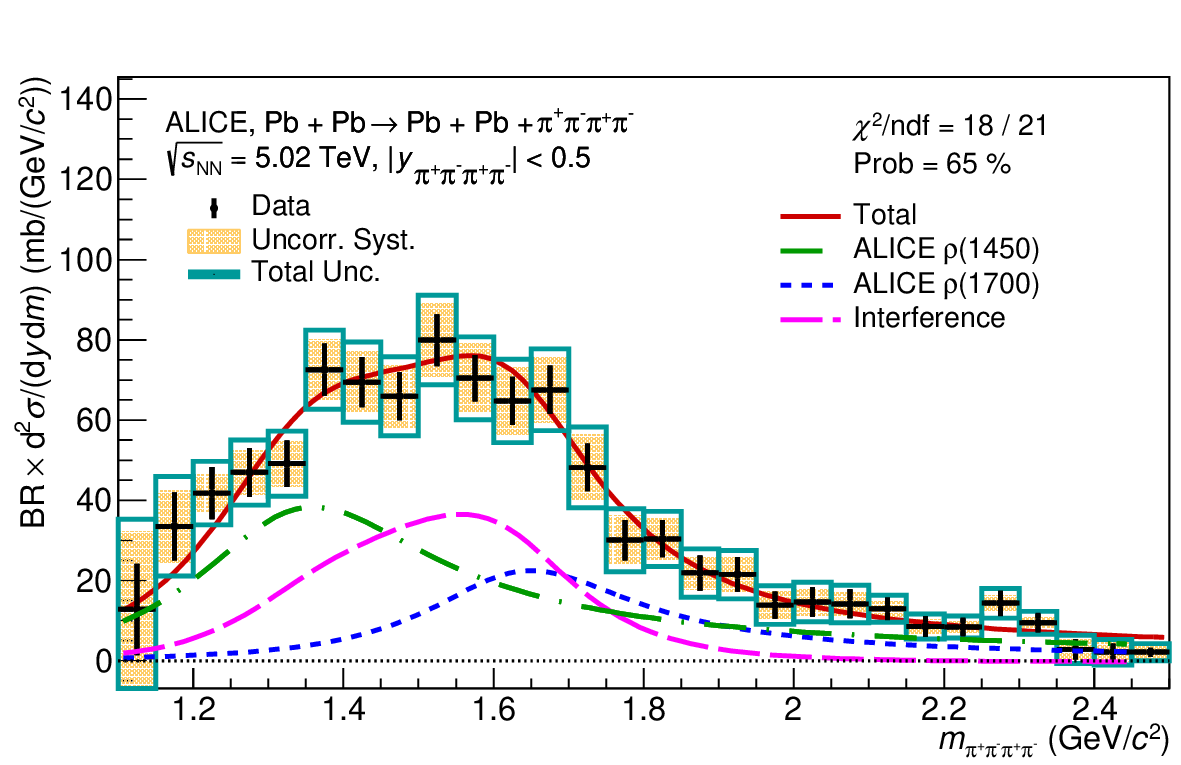}
\caption{\label{sod3} Corrected invariant mass spectrum for the coherent four pion photoproduction fitted with one resonance (top), one resonance interfering with non-resonant production (middle) and two resonance with interference (bottom) models, as described in the text. Black error bars represent the statistical error, the orange band shows the uncorrelated systematic uncertainty, and the boxes show them added in quadrature. "Prob" reflects the probability of having the given or a higher $\chi^2/\rm{ndf}$ of the fit.}
\end{figure}

The panels in Fig.~\ref{sod3} show the results of the Log-Likelihood fits to the fully corrected invariant mass cross sections using the expressions (\ref{SR}),  (\ref{SRinter}) and (\ref{TRinter}), respectively. The total uncorrelated uncertainty (teal open box), calculated as a statistical (black cross) and uncorrelated systematic (filled orange box) added in quadrature, is taken into account in the fits. The results of these fits are presented in Table~\ref{tab:fit}, together with the PDG values and the previous measurement by the STAR Collaboration~\cite{STAR:2009giy} for comparison.

\par  The fit considering one B--W resonance provides mass and width parameters consistent with the $\rho(1450)$ resonance reported by the Particle Data Group~\cite{ParticleDataGroup:2022pth}. This result is 80~\MeVmass, or about $2\sigma$, lower than the value reported by STAR~\cite{STAR:2009giy} for the \fourpi channel.

\par The fits by Eqs.~\ref{SRinter} and~\ref{TRinter} considerably improve the description of the data, especially in the higher-mass region. For the case of two interfering resonances, the mass of the lighter resonance is lower than the one reported by PDG for $\rho(1450)$, but still statistically compatible with it at a 1.7~$\sigma$, taking into account both the statistical and the systematic uncertainties of the current measurement. The mass of the heavier resonance is also slightly lower (2.2~$\sigma$) than the one reported by PDG. The widths of both resonances agree with the corresponding PDG values within the reported uncertainties. For the fit including interference with non-resonant four-pion production (\ref{SRinter}), the fit returns the mass value in the middle between the two resonances reported by PDG for $\rho(1450)$ and $\rho(1700)$, respectively, at 3.0~$\sigma$ from each of them, and the width of the resonance is significantly larger than the PDG values for these resonances. The fit with two interfering resonances provides results that are consistent with the resonances in the PDG. Other parameterizations that rely on alternative models of the non-resonant four-pion production, as shown in Ref.~\cite{H1:2018-011}, might  provide equally good description of the invariant mass distribution for the case of the single broad resonance scenario (see Supplementary Fig. S1). Using alternative forms for the background did not significantly change the extracted mass and width of the single resonance.

\par The probabilities for the single, two resonance and single resonance with interference hypotheses producing $\chi^2$ less than the observed values of $\chi^2$ are 0.4, 65 and 85\%, respectively. The single resonance hypothesis without an interference term is thus strongly disfavored. A more complex fit including two resonances and a non-resonant component was also tested (see Supplementary Fig. S2). It showed no improvement in fit quality, yielded a negligible continuum contribution, and significantly increased parameter uncertainties due to the increased number of free parameters of the fit. For these reasons it is not included in the paper. Notably, it also yielded a negligible continuum contribution which is consistent with the Regge-based calculation performed for the two-resonance scenario~\cite{Kycia:2017iij}.

\begin{table}[h!]
\centering
\caption{Summary of the fit results. The first three rows present the masses and widths reported by PDG~\cite{ParticleDataGroup:2022pth} and measured by the STAR Collaboration~\cite{STAR:2009giy}, while the second and third part list the values extracted from each of the fits by Eq.~\ref{SR}, Eq.~\ref{SRinter} and by Eq.~\ref{TRinter}, respectively. The given uncertainties for the ALICE results are statistical and systematic, respectively. STAR reports only statistical uncertainty, and the uncertainties in the PDG are their best estimate.}
\begin{tabular}{lrr}
\hline
 & $m$ (\MeVmass)  &  $\Gamma$ (\MeVmass)  \\
 \hline
PDG $\rho(1450)$ &  1465 $\pm$ 25 &  400 $\pm$ 60   \\
PDG $\rho(1700)$ &  1720 $\pm$ 20 &  250 $\pm$ 100 \\
STAR Au--Au & 1540 $\pm$ 40 &  570 $\pm$ 60 \\
\hline
ALICE Pb--Pb single resonance &  1463$~\pm$ 2 $\pm$ 15&  448$~\pm$ 6 $\pm$ 14  \\
\hline
ALICE Pb--Pb single resonance \\
with non-resonant continuum  &  1595$~\pm$ 26 $\pm$ 34&  587$~\pm$ 50 $\pm$ 61  \\
\hline
ALICE Pb--Pb $\rho(1450)$&  1385$~\pm$ 14 $\pm$ 36 &  431$~\pm$ 36 $\pm$ 82  \\
ALICE Pb--Pb $\rho(1700)$ & 1663$~\pm$ 13 $\pm$ 22  &  357$~\pm$ 31  $\pm$ 49  \\
Mixing angle & \multicolumn{2}{c}{1.52$~\pm$ 0.16 $\pm$ 0.19 (rad)} \\
\hline
\end{tabular}
\label {tab:fit}
\end{table}

\par
Previous measurements reported by the STAR~\cite{STAR:2009giy} and the ALICE~\cite{ALICE:2020ugp} Collaborations for the two-pion decay channel did not observe any resonance around the $\rho(1450)$ mass. This fact is not a contradiction, but instead can be explained by the hypothesis of the existence of multiple excited $\rho$ states with the two-pion decay channel being strongly suppressed for $\rho(1450)$. 

\section{Systematic uncertainties}
Several sources of systematic uncertainties in the cross section measurement are considered in this analysis. The uncorrelated uncertainties related to the acceptance and trigger efficiency and to background estimation can influence the shape of the invariant mass distribution. They are taken into account before performing the fits to the spectra. They will influence the cross section, the mass, and the width of the resonances extracted. Other sources of systematic uncertainty are correlated across the invariant mass spectra, as they affect only the overall normalization, so they are added quadratically only for the cross section calculations.

\par The largest source of systematic uncertainty is geometric acceptance and trigger efficiency. It originates in the uncertainty on the angular distribution of the \fourpi~final state. Following the STAR measurement~\cite{STAR:2009giy}, where the decay mode \rhopipi was found to be preferred, in this analysis the excited $\rho$ from STARlight was forced to decay according to this mode (which is different from the standard decay mode in STARlight). The subsequent decay of $\rho^0\rightarrow\pi^+\pi^-$ takes into account the spin of the $\rho^0$ being 1. Events generated in this way were then used to calculate the acceptance and efficiency. At the same time, the PDG~\cite{ParticleDataGroup:2022pth} mentions several other decay channels, in particular $\rho\rightarrow\pi^+\pi^-\pi^+\pi^-$ without an intermediate $\rho^0$ resonance production. Various angular distributions between the final state particles of these decay channels would result in different fractions of the four pions to be found inside the tracker acceptance, and thus in the (uncorrelated) variations of the $A\times\varepsilon$ corrections. The second possible effect (correlated) is related to the trigger used in this analysis, which requires a large opening angle between the selected tracks. Therefore, the estimated trigger efficiency can also be affected by the angular distribution in the final state particles. To study these effects, the azimuthal angular distribution between two positive pions (the distribution for two negative pions is identical) in an event is reweighted to match the flat (isotropic) distribution~\cite{STAR:2009giy}. The weight calculated for each event at the generator level is then propagated to estimate the $A\times\varepsilon$ corrections. The correlated part of this uncertainty is 12\%, while the uncorrelated part amounts to 6.5\%.

\par The uncertainty related to the background subtraction is estimated by varying the scale factor used to estimate the contribution from the combinatorial background obtained from the template fit \mbox{($1.60\pm0.15~\rm{(stat.)}$)} within its uncertainty, between 1.45 and 1.75. The effect on the extracted cross section is 1.5\%. The uncertainty associated with the reweighting of the \pt~spectrum of the STARlight MC was found to be negligible.

\par The B--W fits are performed with random combinations of the lower and upper limits of the fit range and of the bin width. These variations result in 1.7\% uncertainty for the total cross section measurement. They also have a dominant effect on the systematic uncertainties of the parameters of the resonances (masses, widths and the mixing angle) extracted from fits. The reported values correspond to the average over all such fits, and the related systematic uncertainty is calculated as the root-mean-square deviation between them.

\par The uncertainty related to the remaining incoherent contribution is estimated by varying the requirement on the total transverse momentum of the \fourpi~final state from 0.1 to 0.2~\GeVc. It is 1.5\%.

\par The uncertainty on the track selection is estimated by changing the required number of TPC clusters matched to the track from 50 to 70 and repeating the complete analysis. The uncertainty corresponds to the full variation of the results and amounts to 1.5\%. The uncertainty on the matching of TPC and ITS tracks is obtained by comparing the behaviour of real and simulated data under different detector conditions and is found to be 4\%~\cite{ALICE:2020ugp}.

\par The uncertainty associated with the determination of the trigger efficiency of the SPD chips is obtained by changing the requirements on the events used for this data-driven method. In real data it could be that the SPD has other signals, e.g. from noise or soft electron-positron production. The performance of the matching algorithm is checked by comparing the results of applying it in data and in MC, where these extra effects are not present. A discrepancy of 1.0\% is found and assigned as a systematic uncertainty. 

\par  The probability of the occurrence of pileup is correlated with the average number of inelastic hadronic collisions per bunch crossing, $\mu$. The uncertainty of the pileup correction is taken as the difference between the two methods used for its calculation~\cite{ALICE:2020ugp}. One method uses an event sample obtained with an unbiased trigger based only on the timing of bunches crossing the interaction region. The second method divides the signal sample into subsets of events with a specific range of $\mu$ values. The uncertainty of this correction is taken as the difference between these two methods
and found to be 3.8\% for the cross section measurement.

\par The uncertainty on the luminosity (2.6\%) has two contributions which were added in quadrature, one from the measurement of the reference cross sections in van der Meer scans (2.5\%~\cite{ALICE:2022xir}) and another from the determination of the live-time of the trigger used in this analysis (0.4\%).

Table~\ref{tab:sys} lists the sources of systematic uncertainties for the extracted cross section. The first two rows present the uncorrelated systematic uncertainties considered in the fits of the invariant mass distributions. The rest of the table shows the correlated systematic uncertainties that influence only the extraction of the cross section.

\begin{table}[h!]
\centering
\caption{Summary of the systematic uncertainties. First two rows show the systematic uncertainties uncorrelated across the invariant mass spectra which are taken into account while performing the fits. The second part of the table presents the sources of the correlated systematic uncertainty and their corresponding total value that is used in the calculations of the cross sections.}
\begin{tabular}{lr}
\hline
 Source & Uncertainty ($\%$)  \\
 \hline
Background subtraction & $1.5$ \\ 
Angular distribution & $6.5$  \\ 
\hline
Total uncorrelated & $6.7$ \\
\hline
Angular distribution  & $12.0$  \\ 
Signal extraction & $1.7$ \\
Track selection & $1.5$ \\
Track matching & $4.0$ \\
Incoherent contribution& $1.5$ \\ 
Trigger efficiency & $1.0$\\
Pileup & $3.8$\\ 
Luminosity & $2.6$  \\
\hline
Total correlated & $13.7$ \\
\hline
\end{tabular}
\label {tab:sys}
\end{table}

\section{Cross section measurement}

\par The total coherent cross section times the branching ratio is obtained by integrating the fitted single B--W distribution over the invariant mass range \mbox{(0.8--2.5~\GeVmass)} and is found to be d$\sigma(|y|<0.5)/$d$y = 47.8\pm2.3~\rm{(stat.)}\pm7.7~\rm{(syst.)}$~mb. The upper limit of the integration at 2.5~\GeVmass~is chosen to avoid the region where the fit significantly overestimates the data, ensuring a more accurate calculation of the cross section. The values obtained from the Eq.~\ref{TRinter} fit by integrating each of the B--W resonances individually over the same mass range are $24.8\pm2.5~\rm{(stat.)}\pm8.1~\rm{(syst.)}$~mb and $10.1\pm2.3~\rm{(stat.)}\pm5.3~\rm{(syst.)}$~mb for $\rho(1450)$ and $\rho(1700)$, respectively. The individual resonances exhibit significantly larger uncertainty values because of the correlation between the two contributions. The sum of these two cross sections is lower than the total observed cross section of the \fourpi~state due to the large interference component.

\par
\begin{figure}[!h]
\includegraphics[width=0.8\textwidth]{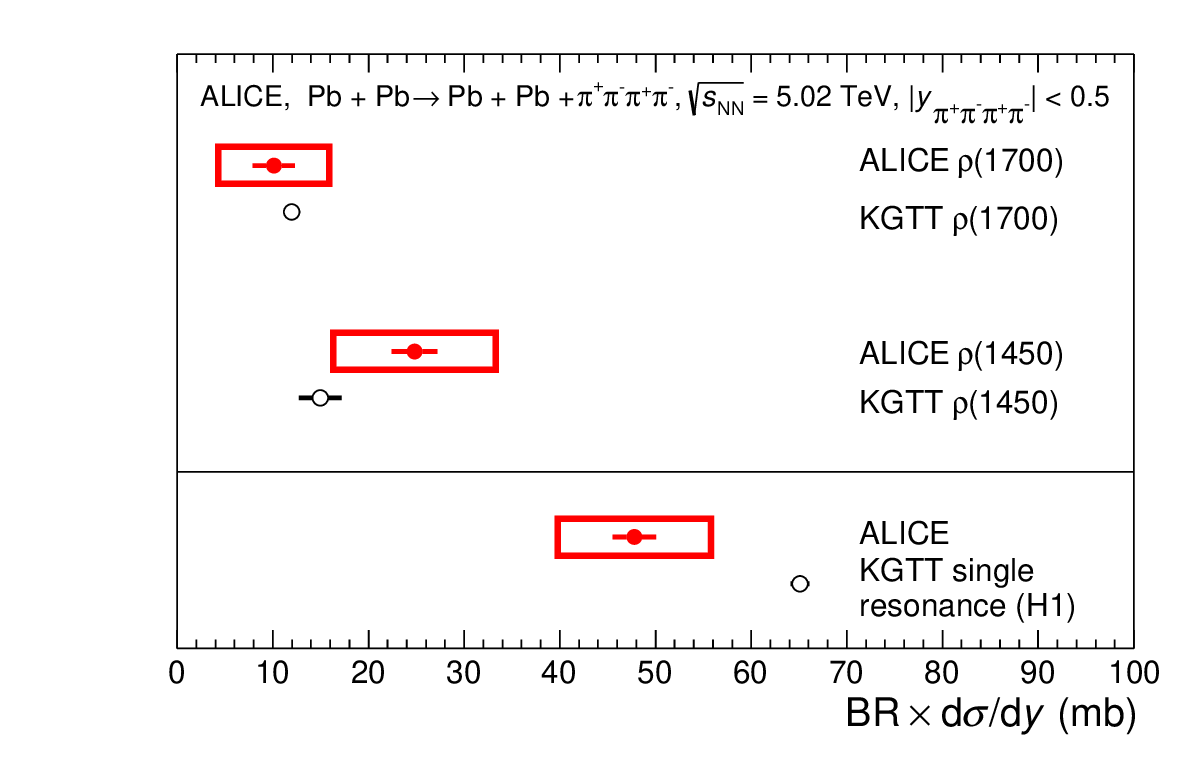}
\caption{\label{mar} Cross sections for the one-resonance and two-resonances scenarios (full red circles) compared to the theoretical calculations from Ref.~\cite{Klusek-Gawenda:2020gwa} (open black circles). Horizontal error bars on the data points show the statistical uncertainty and the boxes represent the statistical and systematic uncertainties added in quadrature. Horizontal error bars on the theory points show the theoretical uncertainty.}
\end{figure}

 The final results are compared to a recent theoretical KGTT model (Klusek-Gawenda and Tapia Takaki)~\cite{Klusek-Gawenda:2020gwa}, as shown in Fig.~\ref{mar}. The KGTT model takes as input a $\gamma-$p cross section based on the Donnachie-Landshoff model, which assumes Reggeon and Pomeron exchange. This cross section is then scaled to the photonuclear interaction using a Glauber model. This model represents two separate calculations: one is done for two excited $\rho$ mesons, $\rho(1450)$ and $\rho(1700)$, and another for a one broad Breit-Wigner resonance to account for the recent H1 results~\cite{H1:2018-011} that suggested this possibility. As for this analysis, the calculations were performed for the excited $\rho$ meson decaying into the $\pi^+ \pi^+ \pi^- \pi^-$ final state. For the case of one broad resonance, the cross section is given for the single resonance component only, without accounting for the non-resonant continuum or interference, therefore only the comparison with the cross section extracted from the Eq.~\ref{SR} fit is presented. Our data is in good agreement with the KGTT model that considers two excited $\rho$ states, while it diverges by $2.1\sigma$ from the KGTT calculation based on a singular resonance.

Additionally, one can compare the ratio between the $\rho\rightarrow \pi^+ \pi^- \pi^+ \pi^-$ and $\rho^0\rightarrow \pi^+ \pi^-$ cross sections, $\mathrm{BR}\times\sigma_{\rho} / \sigma_{\rho^0}$, with the one measured by the STAR Collaboration~\cite{STAR:2009giy}. This is done by dividing the total observed cross section by the results reported recently by the ALICE collaboration in Ref.~\cite{ALICE:2020ugp}. The $\rho^0$ analysis used the same trigger requirements and analysis selections, leading to partial cancellation of the correlated systematic uncertainties (related to trigger efficiency, pileup and luminosity). The obtained value of the $(\rho\rightarrow \pi^+ \pi^- \pi^+ \pi^-)/(\rho^0\rightarrow \pi^+ \pi^-)$ ratio is $0.088 \pm 0.004~\rm{(stat.)} \pm 0.013~\rm{(syst.)}$ for the central rapidity region $|y|<0.5$. One has to extrapolate this ratio to the full solid angle to compare it with the corresponding result from STAR~\cite{STAR:2009giy}. The extrapolation factors were calculated as the ratios of the photoproduction cross sections in the full rapidity interval to the cross section in the measured rapidity region. Their values were obtained using the STARlight MC~\cite{Klein:2016yzr} and the theoretical calculation from Ref.~\cite{Klusek-Gawenda:2020gwa}, with the difference between these two approaches taken as the systematic uncertainty. The extrapolation factors are $8.8\pm0.1~\rm{(syst.)}$ and $10.6\pm0.1~\rm{(syst.)}$ for excited $\rho$ and $\rho^0$, respectively. The resulting cross section ratio, expressed in percent, is presented in Table~\ref{tab:sum}, together with the STAR result. Our measurement is lower than the one reported by the STAR Collaboration which found different single-resonance masses and widths, and carried out their measurements accompanied by forward neutron emission due to mutual nuclear excitation, precluding a comprehensive comparison. Concurrently, the observed reduction in this ratio with increasing collision energies, from RHIC to the LHC, aligns qualitatively with theoretical predictions by the KGTT model~\cite{Klusek-Gawenda:2020gwa}, attributable to a more rapid reduction of Reggeon exchange contributions in excited $\rho$ compared to $\rho^0$ photoproduction.

\begin{table}[h!]
\centering
\caption{Summary of the cross section ratio measurement. The first part presents the cross section values both for the single and two resonance scenarios, as extracted from the fits by Eq.~\ref{SR} and by Eq.~\ref{TRinter}, respectively. The second part lists the ratio between the $\rho\rightarrow \pi^+ \pi^- \pi^+ \pi^-$ and $\rho^0\rightarrow \pi^+ \pi^-$ cross sections expressed in percent obtained in this measurement and also reported by the STAR Collaboration~\cite{STAR:2009giy}. Note that the STAR Collaboration performed this measurement for the case of mutual nuclear excitation. The given uncertainties are statistical and systematic, respectively. }
\begin{tabular}{lcc}
&& $\mathrm{BR} \times \sigma/\mathrm{d}y$ (mb) \\
\hline
ALICE Pb--Pb single resonance && $47.8\pm2.3\pm7.7$\\
ALICE Pb--Pb $\rho(1450)$&&    $24.8\pm2.5\pm8.1$\\
ALICE Pb--Pb $\rho(1700)$ &&   $10.1\pm2.3\pm5.3$\\
\hline
& $\sqrt{s_{\mathrm{NN}}}$ & $\sigma(\rho \rightarrow \pi^+ \pi^- \pi^+ \pi^-)/\sigma(\rho^0 
\rightarrow \pi^+ \pi^- )$  \\
 
 \hline
STAR Au--Au~\cite{STAR:2009giy} & 200~\GeV  & $(13.4\pm0.8\pm 4.4)\%$\\ 
ALICE Pb--Pb & 5.02~\TeV &  $(7.3 \pm 0.4\pm 1.2)\%$\\ 
\hline
\end{tabular}
\label {tab:sum}
\end{table}

\section{Summary}
The coherent \fourpi~production was studied for the first time in ultraperipheral Pb--Pb collisions at the LHC. 
The four-pion cross section integrated over the invariant mass range (0.8--2.5)~\GeVmass~is d$\sigma(|y|<0.5)/$d$y = 47.8\pm2.3~\rm{(stat.)}\pm7.7~\rm{(syst.)}$~mb.
The peak around the invariant mass 1.5~\GeVmass~is consistent with the results reported by STAR Collaboration~\cite{STAR:2009giy}. The \fourpi~invariant mass distribution can be described either by a single resonance plus a constant interference term or by two interfering BW-distributions, $\rho(1450)$ and $\rho(1700)$, and the interference term between them. Different parameterizations of the non-resonant four-pion production contribution have been explored in the literature~\cite{H1:2018-011}.
A cross check has shown that such a non-resonant contribution can give an equally good fit of the data, but it does not significantly modify the extracted mass and width of the resonance. The fit with two interfering resonances gives results in agreement with resonances currently included in the PDG. The extracted masses and widths of the two resonances are $m_{1} = 1385 \pm 14~\rm{(stat.)}\pm 36~\rm{(syst.)}$~\MeVmass~and $\Gamma_{1} = 431~\pm 36~\rm{(stat.)}\pm 82~\rm{(syst.)}$~\MeVmass, and $m_{2} = 1663~\pm 13~\rm{(stat.)}\pm 22~\rm{(syst.)}$~\MeVmass~and $\Gamma_{2} = 357~\pm 31~\rm{(stat.)}\pm 49~\rm{(syst.)}$~\MeVmass, respectively. The mixing angle between the two resonances is $\varphi=1.52~\pm0.16~\rm{(stat.)}\pm0.19~\rm{(syst.)}$~rad.
The extracted cross section values are compared to recent theoretical calculations~\cite{Klusek-Gawenda:2020gwa}. 
The ratio of the cross sections of $\rho$ to $\rho^0$ was also studied and is somewhat lower than the one measured by STAR in the events with mutual nuclear excitation. The observed reduction in this ratio with increasing collision energies could also be attributed to a more rapid reduction of Reggeon exchange contributions in excited $\rho$ compared to $\rho^0$ photoproduction.

\newenvironment{acknowledgement}{\relax}{\relax}
\begin{acknowledgement}
\section*{Acknowledgements}
% Version: 2024-03-22

The ALICE Collaboration would like to thank all its engineers and technicians for their invaluable contributions to the construction of the experiment and the CERN accelerator teams for the outstanding performance of the LHC complex.
The ALICE Collaboration gratefully acknowledges the resources and support provided by all Grid centres and the Worldwide LHC Computing Grid (WLCG) collaboration.
The ALICE Collaboration acknowledges the following funding agencies for their support in building and running the ALICE detector:
A. I. Alikhanyan National Science Laboratory (Yerevan Physics Institute) Foundation (ANSL), State Committee of Science and World Federation of Scientists (WFS), Armenia;
Austrian Academy of Sciences, Austrian Science Fund (FWF): [M 2467-N36] and Nationalstiftung f\"{u}r Forschung, Technologie und Entwicklung, Austria;
Ministry of Communications and High Technologies, National Nuclear Research Center, Azerbaijan;
Conselho Nacional de Desenvolvimento Cient\'{\i}fico e Tecnol\'{o}gico (CNPq), Financiadora de Estudos e Projetos (Finep), Funda\c{c}\~{a}o de Amparo \`{a} Pesquisa do Estado de S\~{a}o Paulo (FAPESP) and Universidade Federal do Rio Grande do Sul (UFRGS), Brazil;
Bulgarian Ministry of Education and Science, within the National Roadmap for Research Infrastructures 2020-2027 (object CERN), Bulgaria;
Ministry of Education of China (MOEC) , Ministry of Science \& Technology of China (MSTC) and National Natural Science Foundation of China (NSFC), China;
Ministry of Science and Education and Croatian Science Foundation, Croatia;
Centro de Aplicaciones Tecnol\'{o}gicas y Desarrollo Nuclear (CEADEN), Cubaenerg\'{\i}a, Cuba;
Ministry of Education, Youth and Sports of the Czech Republic, Czech Republic;
The Danish Council for Independent Research | Natural Sciences, the VILLUM FONDEN and Danish National Research Foundation (DNRF), Denmark;
Helsinki Institute of Physics (HIP), Finland;
Commissariat \`{a} l'Energie Atomique (CEA) and Institut National de Physique Nucl\'{e}aire et de Physique des Particules (IN2P3) and Centre National de la Recherche Scientifique (CNRS), France;
Bundesministerium f\"{u}r Bildung und Forschung (BMBF) and GSI Helmholtzzentrum f\"{u}r Schwerionenforschung GmbH, Germany;
General Secretariat for Research and Technology, Ministry of Education, Research and Religions, Greece;
National Research, Development and Innovation Office, Hungary;
Department of Atomic Energy Government of India (DAE), Department of Science and Technology, Government of India (DST), University Grants Commission, Government of India (UGC) and Council of Scientific and Industrial Research (CSIR), India;
National Research and Innovation Agency - BRIN, Indonesia;
Istituto Nazionale di Fisica Nucleare (INFN), Italy;
Japanese Ministry of Education, Culture, Sports, Science and Technology (MEXT) and Japan Society for the Promotion of Science (JSPS) KAKENHI, Japan;
Consejo Nacional de Ciencia (CONACYT) y Tecnolog\'{i}a, through Fondo de Cooperaci\'{o}n Internacional en Ciencia y Tecnolog\'{i}a (FONCICYT) and Direcci\'{o}n General de Asuntos del Personal Academico (DGAPA), Mexico;
Nederlandse Organisatie voor Wetenschappelijk Onderzoek (NWO), Netherlands;
The Research Council of Norway, Norway;
Pontificia Universidad Cat\'{o}lica del Per\'{u}, Peru;
Ministry of Education and Science, National Science Centre and WUT ID-UB, Poland;
Korea Institute of Science and Technology Information and National Research Foundation of Korea (NRF), Republic of Korea;
Ministry of Education and Scientific Research, Institute of Atomic Physics, Ministry of Research and Innovation and Institute of Atomic Physics and Universitatea Nationala de Stiinta si Tehnologie Politehnica Bucuresti, Romania;
Ministry of Education, Science, Research and Sport of the Slovak Republic, Slovakia;
National Research Foundation of South Africa, South Africa;
Swedish Research Council (VR) and Knut \& Alice Wallenberg Foundation (KAW), Sweden;
European Organization for Nuclear Research, Switzerland;
Suranaree University of Technology (SUT), National Science and Technology Development Agency (NSTDA) and National Science, Research and Innovation Fund (NSRF via PMU-B B05F650021), Thailand;
Turkish Energy, Nuclear and Mineral Research Agency (TENMAK), Turkey;
National Academy of  Sciences of Ukraine, Ukraine;
Science and Technology Facilities Council (STFC), United Kingdom;
National Science Foundation of the United States of America (NSF) and United States Department of Energy, Office of Nuclear Physics (DOE NP), United States of America.
In addition, individual groups or members have received support from:
Czech Science Foundation (grant no. 23-07499S), Czech Republic;
European Research Council (grant no. 950692), European Union;
ICSC - Centro Nazionale di Ricerca in High Performance Computing, Big Data and Quantum Computing, European Union - NextGenerationEU;
Academy of Finland (Center of Excellence in Quark Matter) (grant nos. 346327, 346328), Finland.
    %%%%%%% done by webmaster team
\end{acknowledgement}

\bibliographystyle{utphys}
\bibliography{references}

%\end{thebibliography}

%%%%%%%%% appendix with author list
\newpage
\appendix

\section{Supplementary Material}

\begin{figure}[!h]
  \centering
  \includegraphics[width=.95\linewidth]{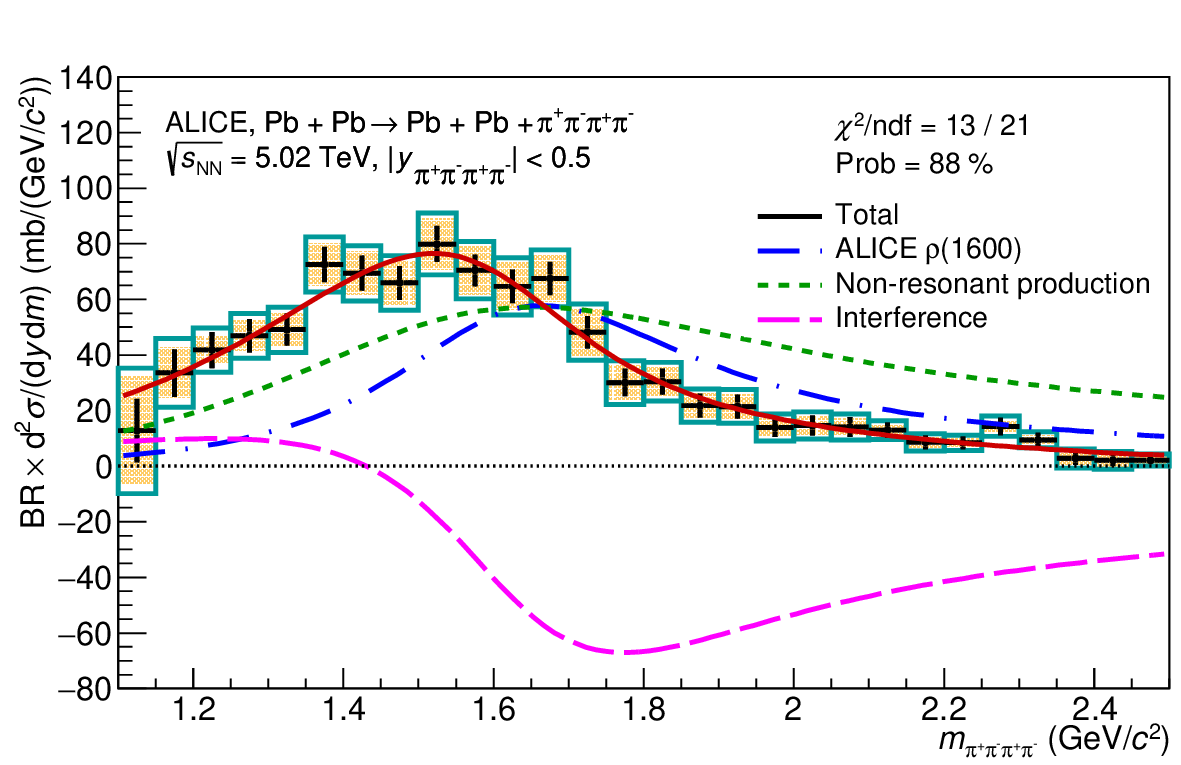}
  \caption{Corrected invariant mass spectrum for the coherent four-pion photoproduction fitted to the model described by a single resonance and an alternative shape of the non-resonant production given by eq.~\ref{eq:supp_Fnr}. Black error bars represent the statistical uncertainty, the orange band shows the uncorrelated systematic uncertainty, and the boxes show them added in quadrature. ``Prob'' reflects the probability of obtaining the given or a higher $\chi^2/\mathrm{ndf}$ of the fit.}
  \label{fig:S2d}
\end{figure}

The following alternative fits to the invariant mass distribution have also been investigated. Figure~\ref{fig:S2d} shows the fit with the parameterization that relies on an alternative model of the non-resonant four-pion production, inspired by Ref.~\cite{H1:2018-011}, which can be expressed by the following equation:
\begin{equation}
  \frac{{\mathrm d}\sigma}{{\mathrm d}m}
  \;=\;
  |A\cdot BW_{1}
  \;+\;
  \,B\cdot \sqrt{F_\mathrm{nr}}\cdot e^{\,i\varphi}
  |^{2}.
  \label{eq:supp_single_expanded}
\end{equation}
Here $BW_{1}$ is the relativistic B--W amplitude defined in the main manuscript, and $A$ is its real amplitude. The parameter $B$ sets the amplitude of the non-resonant continuum, and its shape is given by:
\begin{equation}
  F_\mathrm{nr} \;=\;
  \Theta\!\bigl(m_\mathrm{event}-4m_\pi\bigr)\;
  \frac{\bigl(m_\mathrm{event}^{2}-16\,m_\pi^{2}\bigr)}
       {1+\bigl[(m_\mathrm{event}-M_{0})/\Gamma_{0}\bigr]^{2}}\!,
  \label{eq:supp_Fnr}
\end{equation}
with $M_{0}$ and $\Gamma_{0}$ treated as free parameters of the fit. Here $\Theta(x)$ denotes the unit (Heaviside) step: $\Theta(x)=1$ for $x>0$ and $\Theta(x)=0$ for $x\le 0$, i.e. below the kinematic boundary.

Equation~(\ref{eq:supp_single_expanded}) gives a fit quality comparable to the flat-continuum function presented in the main manuscript ($\chi^{2}/\mathrm{ndf}=13/21$ vs.\ $15/22$) and returns $m=1686\pm95~\mathrm{(stat.)}$~\MeVmass\ and $\Gamma=543\pm127~\mathrm{(stat.)}$~\MeVmass, consistent within uncertainties with the single-resonance and a flat continuum model, consistent within uncertainties with the single-resonance and a flat continuum model, but with considerably larger uncertainties. 

\vspace{1em}
% ---------- S2 ----------
In addition, the non-resonant four-pion production along with the two interfering resonances is also considered:
\begin{equation}
  \frac{{\mathrm d}\sigma}{{\mathrm d}m}
  \;=\;
  \left|\,A\cdot BW_{1}
        + e^{-i\varphi}\cdot B\cdot BW_{2}
        + e^{-i\psi}\cdot C\,\right|^{2},
  \label{eq:supp_two_coherent}
\end{equation}
where the amplitudes, masses, and widths of the resonances and the two mixing angles are treated as free parameters of the fit.

\begin{figure}[!h]
  \centering
  \includegraphics[width=.95\linewidth]{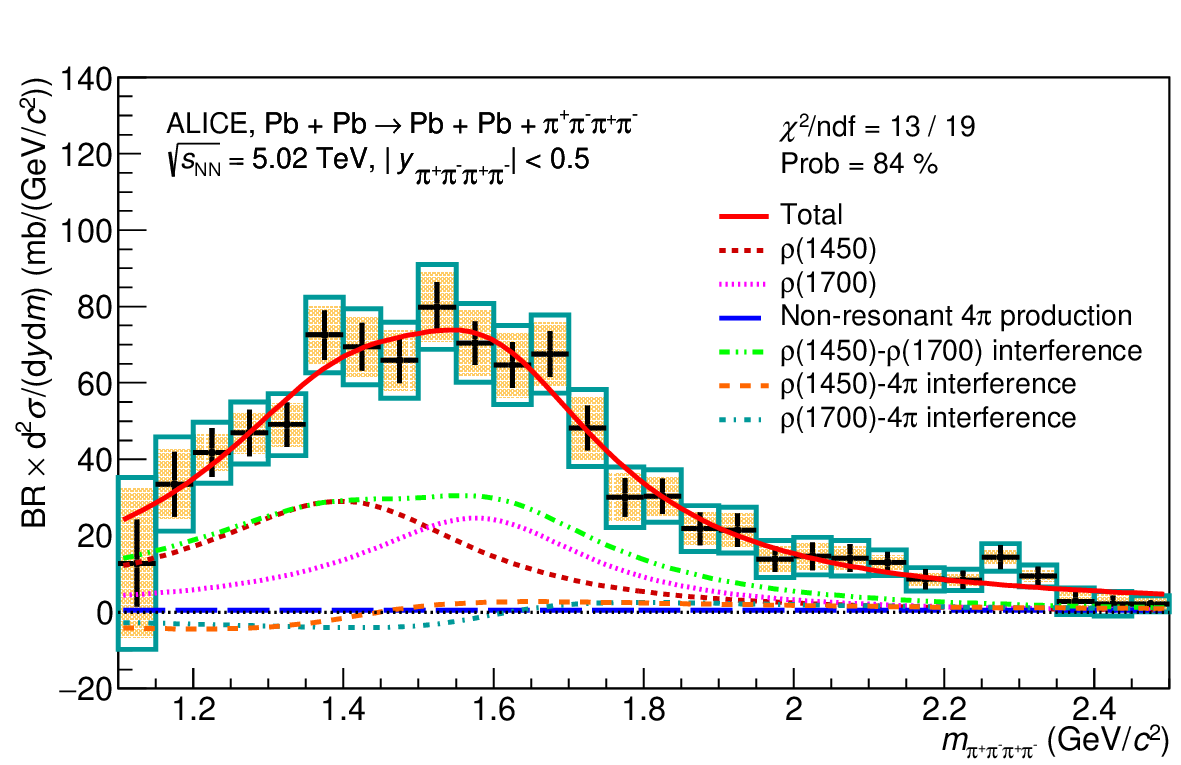}
  \caption{Corrected invariant mass spectrum for the coherent four-pion photoproduction fitted with two resonances plus a non-resonant continuum, eq.~\ref{eq:supp_two_coherent}. Black error bars represent the statistical uncertainty, the orange band shows the uncorrelated systematic uncertainty, and the boxes show them added in quadrature. ``Prob'' reflects the probability of obtaining the given or a higher $\chi^2/\mathrm{ndf}$ of the fit}
  \label{fig:S2e}
\end{figure}

Adding the coherent constant non-resonant term in Eq.~(\ref{eq:supp_two_coherent}) slightly changes $\chi^{2}/\mathrm{ndf}$ from $18/21$ to $13/19$, with masses and widths compatible with Table~1 of the manuscript, while the continuum normalization is consistent with zero. The additional freedom increases the uncertainties of all extracted values without a decisive gain in description quality; this variant is therefore documented only in this supplementary material.

               %%%%%%%%%%% put your appendices here
\clearpage
\section{The ALICE Collaboration}
\label{app:collab}
% ALICE Collaboration author list for 2024-03-22
\begin{flushleft} 
\small

S.~Acharya\,\orcidlink{0000-0002-9213-5329}\,$^{\rm 127}$, 
D.~Adamov\'{a}\,\orcidlink{0000-0002-0504-7428}\,$^{\rm 86}$, 
A.~Agarwal$^{\rm 135}$, 
G.~Aglieri Rinella\,\orcidlink{0000-0002-9611-3696}\,$^{\rm 32}$, 
L.~Aglietta\,\orcidlink{0009-0003-0763-6802}\,$^{\rm 24}$, 
M.~Agnello\,\orcidlink{0000-0002-0760-5075}\,$^{\rm 29}$, 
N.~Agrawal\,\orcidlink{0000-0003-0348-9836}\,$^{\rm 25}$, 
Z.~Ahammed\,\orcidlink{0000-0001-5241-7412}\,$^{\rm 135}$, 
S.~Ahmad\,\orcidlink{0000-0003-0497-5705}\,$^{\rm 15}$, 
S.U.~Ahn\,\orcidlink{0000-0001-8847-489X}\,$^{\rm 71}$
% ...

\section*{Affiliation Notes}

$^{\rm I}$ Deceased\\
$^{\rm II}$ Also at: Max-Planck-Institut fur Physik, Munich, Germany\\
$^{\rm III}$ Also at: Italian National Agency for New Technologies, Energy and Sustainable Economic Development (ENEA), Bologna, Italy\\
$^{\rm IV}$ Also at: Dipartimento DET del Politecnico di Torino, Turin, Italy\\
$^{\rm V}$ Also at: Yildiz Technical University, Istanbul, T\"{u}rkiye\\
$^{\rm VI}$ Also at: Department of Applied Physics, Aligarh Muslim University, Aligarh, India\\
$^{\rm VII}$ Also at: Institute of Theoretical Physics, University of Wroclaw, Poland\\
$^{\rm VIII}$ Also at: An institution covered by a cooperation agreement with CERN\\

\section*{Collaboration Institutes}

$^{1}$ A.I. Alikhanyan National Science Laboratory (Yerevan Physics Institute) Foundation, Yerevan, Armenia\\
$^{2}$ AGH University of Krakow, Cracow, Poland\\
$^{3}$ Bogolyubov Institute for Theoretical Physics, National Academy of Sciences of Ukraine, Kiev, Ukraine\\
$^{4}$ Bose Institute, Department of Physics  and Centre for Astroparticle Physics and Space Science (CAPSS), Kolkata, India\\
$^{5}$ California Polytechnic State University, San Luis Obispo, California, United States\\
$^{6}$ Central China Normal University, Wuhan, China\\
$^{7}$ Centro de Aplicaciones Tecnol\'{o}gicas y Desarrollo Nuclear (CEADEN), Havana, Cuba\\
$^{8}$ Centro de Investigaci\'{o}n y de Estudios Avanzados (CINVESTAV), Mexico City and M\'{e}rida, Mexico\\
$^{9}$ Chicago State University, Chicago, Illinois, United States\\
$^{10}$ China Institute of Atomic Energy, Beijing, China\\
$^{11}$ China University of Geosciences, Wuhan, China\\
$^{12}$ Chungbuk National University, Cheongju, Republic of Korea\\
$^{13}$ Comenius University Bratislava, Faculty of Mathematics, Physics and Informatics, Bratislava, Slovak Republic\\
$^{14}$ Creighton University, Omaha, Nebraska, United States\\
$^{15}$ Department of Physics, Aligarh Muslim University, Aligarh, India\\
$^{16}$ Department of Physics, Pusan National University, Pusan, Republic of Korea\\
$^{17}$ Department of Physics, Sejong University, Seoul, Republic of Korea\\
$^{18}$ Department of Physics, University of California, Berkeley, California, United States\\
$^{19}$ Department of Physics, University of Oslo, Oslo, Norway\\
$^{20}$ Department of Physics and Technology, University of Bergen, Bergen, Norway\\
$^{21}$ Dipartimento di Fisica, Universit\`{a} di Pavia, Pavia, Italy\\
$^{22}$ Dipartimento di Fisica dell'Universit\`{a} and Sezione INFN, Cagliari, Italy\\
$^{23}$ Dipartimento di Fisica dell'Universit\`{a} and Sezione INFN, Trieste, Italy\\
$^{24}$ Dipartimento di Fisica dell'Universit\`{a} and Sezione INFN, Turin, Italy\\
$^{25}$ Dipartimento di Fisica e Astronomia dell'Universit\`{a} and Sezione INFN, Bologna, Italy\\
$^{26}$ Dipartimento di Fisica e Astronomia dell'Universit\`{a} and Sezione INFN, Catania, Italy\\
$^{27}$ Dipartimento di Fisica e Astronomia dell'Universit\`{a} and Sezione INFN, Padova, Italy\\
$^{28}$ Dipartimento di Fisica `E.R.~Caianiello' dell'Universit\`{a} and Gruppo Collegato INFN, Salerno, Italy\\
$^{29}$ Dipartimento DISAT del Politecnico and Sezione INFN, Turin, Italy\\
$^{30}$ Dipartimento di Scienze MIFT, Universit\`{a} di Messina, Messina, Italy\\
$^{31}$ Dipartimento Interateneo di Fisica `M.~Merlin' and Sezione INFN, Bari, Italy\\
$^{32}$ European Organization for Nuclear Research (CERN), Geneva, Switzerland\\
$^{33}$ Faculty of Electrical Engineering, Mechanical Engineering and Naval Architecture, University of Split, Split, Croatia\\
$^{34}$ Faculty of Engineering and Science, Western Norway University of Applied Sciences, Bergen, Norway\\
$^{35}$ Faculty of Nuclear Sciences and Physical Engineering, Czech Technical University in Prague, Prague, Czech Republic\\
$^{36}$ Faculty of Physics, Sofia University, Sofia, Bulgaria\\
$^{37}$ Faculty of Science, P.J.~\v{S}af\'{a}rik University, Ko\v{s}ice, Slovak Republic\\
$^{38}$ Frankfurt Institute for Advanced Studies, Johann Wolfgang Goethe-Universit\"{a}t Frankfurt, Frankfurt, Germany\\
$^{39}$ Fudan University, Shanghai, China\\
$^{40}$ Gangneung-Wonju National University, Gangneung, Republic of Korea\\
$^{41}$ Gauhati University, Department of Physics, Guwahati, India\\
$^{42}$ Helmholtz-Institut f\"{u}r Strahlen- und Kernphysik, Rheinische Friedrich-Wilhelms-Universit\"{a}t Bonn, Bonn, Germany\\
$^{43}$ Helsinki Institute of Physics (HIP), Helsinki, Finland\\
$^{44}$ High Energy Physics Group,  Universidad Aut\'{o}noma de Puebla, Puebla, Mexico\\
$^{45}$ Horia Hulubei National Institute of Physics and Nuclear Engineering, Bucharest, Romania\\
$^{46}$ HUN-REN Wigner Research Centre for Physics, Budapest, Hungary\\
$^{47}$ Indian Institute of Technology Bombay (IIT), Mumbai, India\\
$^{48}$ Indian Institute of Technology Indore, Indore, India\\
$^{49}$ INFN, Laboratori Nazionali di Frascati, Frascati, Italy\\
$^{50}$ INFN, Sezione di Bari, Bari, Italy\\
$^{51}$ INFN, Sezione di Bologna, Bologna, Italy\\
$^{52}$ INFN, Sezione di Cagliari, Cagliari, Italy\\
$^{53}$ INFN, Sezione di Catania, Catania, Italy\\
$^{54}$ INFN, Sezione di Padova, Padova, Italy\\
$^{55}$ INFN, Sezione di Pavia, Pavia, Italy\\
$^{56}$ INFN, Sezione di Torino, Turin, Italy\\
$^{57}$ INFN, Sezione di Trieste, Trieste, Italy\\
$^{58}$ Inha University, Incheon, Republic of Korea\\
$^{59}$ Institute for Gravitational and Subatomic Physics (GRASP), Utrecht University/Nikhef, Utrecht, Netherlands\\
$^{60}$ Institute of Experimental Physics, Slovak Academy of Sciences, Ko\v{s}ice, Slovak Republic\\
$^{61}$ Institute of Physics, Homi Bhabha National Institute, Bhubaneswar, India\\
$^{62}$ Institute of Physics of the Czech Academy of Sciences, Prague, Czech Republic\\
$^{63}$ Institute of Space Science (ISS), Bucharest, Romania\\
$^{64}$ Institut f\"{u}r Kernphysik, Johann Wolfgang Goethe-Universit\"{a}t Frankfurt, Frankfurt, Germany\\
$^{65}$ Instituto de Ciencias Nucleares, Universidad Nacional Aut\'{o}noma de M\'{e}xico, Mexico City, Mexico\\
$^{66}$ Instituto de F\'{i}sica, Universidade Federal do Rio Grande do Sul (UFRGS), Porto Alegre, Brazil\\
$^{67}$ Instituto de F\'{\i}sica, Universidad Nacional Aut\'{o}noma de M\'{e}xico, Mexico City, Mexico\\
$^{68}$ iThemba LABS, National Research Foundation, Somerset West, South Africa\\
$^{69}$ Jeonbuk National University, Jeonju, Republic of Korea\\
$^{70}$ Johann-Wolfgang-Goethe Universit\"{a}t Frankfurt Institut f\"{u}r Informatik, Fachbereich Informatik und Mathematik, Frankfurt, Germany\\
$^{71}$ Korea Institute of Science and Technology Information, Daejeon, Republic of Korea\\
$^{72}$ KTO Karatay University, Konya, Turkey\\
$^{73}$ Laboratoire de Physique Subatomique et de Cosmologie, Universit\'{e} Grenoble-Alpes, CNRS-IN2P3, Grenoble, France\\
$^{74}$ Lawrence Berkeley National Laboratory, Berkeley, California, United States\\
$^{75}$ Lund University Department of Physics, Division of Particle Physics, Lund, Sweden\\
$^{76}$ Nagasaki Institute of Applied Science, Nagasaki, Japan\\
$^{77}$ Nara Women{'}s University (NWU), Nara, Japan\\
$^{78}$ National and Kapodistrian University of Athens, School of Science, Department of Physics , Athens, Greece\\
$^{79}$ National Centre for Nuclear Research, Warsaw, Poland\\
$^{80}$ National Institute of Science Education and Research, Homi Bhabha National Institute, Jatni, India\\
$^{81}$ National Nuclear Research Center, Baku, Azerbaijan\\
$^{82}$ National Research and Innovation Agency - BRIN, Jakarta, Indonesia\\
$^{83}$ Niels Bohr Institute, University of Copenhagen, Copenhagen, Denmark\\
$^{84}$ Nikhef, National institute for subatomic physics, Amsterdam, Netherlands\\
$^{85}$ Nuclear Physics Group, STFC Daresbury Laboratory, Daresbury, United Kingdom\\
$^{86}$ Nuclear Physics Institute of the Czech Academy of Sciences, Husinec-\v{R}e\v{z}, Czech Republic\\
$^{87}$ Oak Ridge National Laboratory, Oak Ridge, Tennessee, United States\\
$^{88}$ Ohio State University, Columbus, Ohio, United States\\
$^{89}$ Physics department, Faculty of science, University of Zagreb, Zagreb, Croatia\\
$^{90}$ Physics Department, Panjab University, Chandigarh, India\\
$^{91}$ Physics Department, University of Jammu, Jammu, India\\
$^{92}$ Physics Program and International Institute for Sustainability with Knotted Chiral Meta Matter (SKCM2), Hiroshima University, Hiroshima, Japan\\
$^{93}$ Physikalisches Institut, Eberhard-Karls-Universit\"{a}t T\"{u}bingen, T\"{u}bingen, Germany\\
$^{94}$ Physikalisches Institut, Ruprecht-Karls-Universit\"{a}t Heidelberg, Heidelberg, Germany\\
$^{95}$ Physik Department, Technische Universit\"{a}t M\"{u}nchen, Munich, Germany\\
$^{96}$ Politecnico di Bari and Sezione INFN, Bari, Italy\\
$^{97}$ Research Division and ExtreMe Matter Institute EMMI, GSI Helmholtzzentrum f\"ur Schwerionenforschung GmbH, Darmstadt, Germany\\
$^{98}$ Saga University, Saga, Japan\\
$^{99}$ Saha Institute of Nuclear Physics, Homi Bhabha National Institute, Kolkata, India\\
$^{100}$ School of Physics and Astronomy, University of Birmingham, Birmingham, United Kingdom\\
$^{101}$ Secci\'{o}n F\'{\i}sica, Departamento de Ciencias, Pontificia Universidad Cat\'{o}lica del Per\'{u}, Lima, Peru\\
$^{102}$ Stefan Meyer Institut f\"{u}r Subatomare Physik (SMI), Vienna, Austria\\
$^{103}$ SUBATECH, IMT Atlantique, Nantes Universit\'{e}, CNRS-IN2P3, Nantes, France\\
$^{104}$ Sungkyunkwan University, Suwon City, Republic of Korea\\
$^{105}$ Suranaree University of Technology, Nakhon Ratchasima, Thailand\\
$^{106}$ Technical University of Ko\v{s}ice, Ko\v{s}ice, Slovak Republic\\
$^{107}$ The Henryk Niewodniczanski Institute of Nuclear Physics, Polish Academy of Sciences, Cracow, Poland\\
$^{108}$ The University of Texas at Austin, Austin, Texas, United States\\
$^{109}$ Universidad Aut\'{o}noma de Sinaloa, Culiac\'{a}n, Mexico\\
$^{110}$ Universidade de S\~{a}o Paulo (USP), S\~{a}o Paulo, Brazil\\
$^{111}$ Universidade Estadual de Campinas (UNICAMP), Campinas, Brazil\\
$^{112}$ Universidade Federal do ABC, Santo Andre, Brazil\\
$^{113}$ Universitatea Nationala de Stiinta si Tehnologie Politehnica Bucuresti, Bucharest, Romania\\
$^{114}$ University of Cape Town, Cape Town, South Africa\\
$^{115}$ University of Derby, Derby, United Kingdom\\
$^{116}$ University of Houston, Houston, Texas, United States\\
$^{117}$ University of Jyv\"{a}skyl\"{a}, Jyv\"{a}skyl\"{a}, Finland\\
$^{118}$ University of Kansas, Lawrence, Kansas, United States\\
$^{119}$ University of Liverpool, Liverpool, United Kingdom\\
$^{120}$ University of Science and Technology of China, Hefei, China\\
$^{121}$ University of South-Eastern Norway, Kongsberg, Norway\\
$^{122}$ University of Tennessee, Knoxville, Tennessee, United States\\
$^{123}$ University of the Witwatersrand, Johannesburg, South Africa\\
$^{124}$ University of Tokyo, Tokyo, Japan\\
$^{125}$ University of Tsukuba, Tsukuba, Japan\\
$^{126}$ Universit\"{a}t M\"{u}nster, Institut f\"{u}r Kernphysik, M\"{u}nster, Germany\\
$^{127}$ Universit\'{e} Clermont Auvergne, CNRS/IN2P3, LPC, Clermont-Ferrand, France\\
$^{128}$ Universit\'{e} de Lyon, CNRS/IN2P3, Institut de Physique des 2 Infinis de Lyon, Lyon, France\\
$^{129}$ Universit\'{e} de Strasbourg, CNRS, IPHC UMR 7178, F-67000 Strasbourg, France, Strasbourg, France\\
$^{130}$ Universit\'{e} Paris-Saclay, Centre d'Etudes de Saclay (CEA), IRFU, D\'{e}partment de Physique Nucl\'{e}aire (DPhN), Saclay, France\\
$^{131}$ Universit\'{e}  Paris-Saclay, CNRS/IN2P3, IJCLab, Orsay, France\\
$^{132}$ Universit\`{a} degli Studi di Foggia, Foggia, Italy\\
$^{133}$ Universit\`{a} del Piemonte Orientale, Vercelli, Italy\\
$^{134}$ Universit\`{a} di Brescia, Brescia, Italy\\
$^{135}$ Variable Energy Cyclotron Centre, Homi Bhabha National Institute, Kolkata, India\\
$^{136}$ Warsaw University of Technology, Warsaw, Poland\\
$^{137}$ Wayne State University, Detroit, Michigan, United States\\
$^{138}$ Yale University, New Haven, Connecticut, United States\\
$^{139}$ Yonsei University, Seoul, Republic of Korea\\
$^{140}$  Zentrum  f\"{u}r Technologie und Transfer (ZTT), Worms, Germany\\
$^{141}$ Affiliated with an institute covered by a cooperation agreement with CERN\\
$^{142}$ Affiliated with an international laboratory covered by a cooperation agreement with CERN.\\

\end{flushleft} 
  %%%%%%% done by webmaster team

\end{document}